 \documentstyle[12pt]{article}
\begin{document}
 \newcommand{\be}[1]{\begin{equation}\label{#1}}
 \newcommand{\ee}{\end{equation}}
 \newcommand{\beqn}[1]{\begin{eqnarray}\label{#1}}
 \newcommand{\eeqn}{\end{eqnarray}}
\newcommand{\mat}[4]{\left(\begin{array}{cc}{#1}&{#2}\\{#3}&{#4}\end{array}
\right)}
 \newcommand{\matr}[9]{\left(\begin{array}{ccc}{#1}&{#2}&{#3}\\{#4}&{#5}&{#6}\\
{#7}&{#8}&{#9}\end{array}\right)}
 \newcommand{\eps}{\varepsilon}
 \renewcommand{\thefootnote}{\fnsymbol{footnote}}
\begin{flushright}
INFN-FE-13-93 \\
October 1993
\end{flushright}
\vspace{15mm}

 \begin{center}
 {\large\bf Grand Unification of Fermion Masses} \footnote{On the basis of
talks given at the XVI International Warsaw Meeting on Elementary Particle
Physics "New Physics with New Experiments",
Kazimierz, Poland, 24-28 May 1993, and at the II Gran Sasso Summer Institute
"From Particle Physics to Cosmology", Gran Sasso National Laboratory, Italy,
6-17 September 1993.} \\
 \vspace{7mm}
{\large  Zurab G. Berezhiani}\footnote{E-mail: BEREZHIANI@FERRARA.INFN.IT,
39967::BEREZHIANI}\\ [5mm]
{\em Istituto Nazionale di Fisica Nucleare, Sezione di Ferrara, 44100 Ferrara,
Italy, \\ [2mm]
and\\ [2mm]
Institute of Physics, Georgian Academy
of Sciences, 380077 Tbilisi, Georgia}\\ [8mm]
 \end{center}
\vspace{2mm}
\begin{abstract}
After a brief review of the flavour problem we present a new predictive
framework based on SUSY $SO(10)$ theory, where the first family plays
a role of the mass unification point. The inter-family hierarchy is first
generated in a sector of superheavy fermions and then transfered in an
inverse way to ordinary
quarks and leptons by means of the universal seesaw mechanism.
The obtained mass matrices are
simply parametrized by two small coefficients
which can be given by the ratio of the GUT and superstring compactification
scales. The model allows a natural (without fine tuning) doublet-triplet
splitting. It has a strong predictive power,
though no special texture is utilized
in contrast to the known predictive frameworks.
Namely, the model implies that $m_b=4-5$ GeV, $m_s=100-150$ MeV,
$m_u/m_d=0.5-0.7$ and $\tan\!\beta<1.1$.
The Top quark is naturally in the 100 GeV range, but not too heavy:
$m_t<150$ GeV. All CKM mixing angles are in correct range.
The Higgsino mediated $d=5$ operators for the proton decay are naturally
suppressed.

\end{abstract}
\renewcommand{\thefootnote}{\arabic{footnote})}
\setcounter{footnote}{0}

\newpage


{\bf 1. Introduction to  Family Puzzles}
\vspace{3mm}

The Standard Model (SM) is internally consistent from the field theoretical
view, and has been extremely successfull in describing various
experimental data accumulated over the past several years. This suggests that
at presently available energies the SM is literally correct in all its sectors.
It is widely believed, however, that there should be a more fundamental
theory valid at some higher energies. The most important issues that motivate
such a belief include the unification of gauge couplings,
problem of gauge hierarchies and problem of fermion flavours (or families).

In SM all the observed fermions are accomodated in a consistent way.
Three families share the same quantum numbers under
the $SU(3)_C\otimes SU(2)_L\otimes U(1)_Y$ gauge symmetry.
Each family is considered as an anomaly free set of initially massless
chiral fermions. They become massive due to the same Higgs field that gives
masses to $W^{\pm}$ and Z bosons. An important feature of the minimal SM is
that the flavour changing neutral currents (FCNC) are naturally suppressed
in both gauge and Higgs boson exchanges \cite{FCNC}.
However, the pattern of fermion masses and mixing remains undetermined
due to arbitrariness of the Yukawa couplings. A hypothetical fundamental
theory should allow to calculate these couplings, or at least somehow
constrain them. When thinking of such a theory, one should bear in mind
that the flavour problem has different aspects, questioning the
origin of (i) family replication  (ii) mass and mixing pattern of charged
fermions (iii) CP violation in weak interactions
(iv) CP conservation in strong interactions (v) tiny neutrino masses.

There is an almost holy trust that all the fundamental problems, and among
them the problem of fermion flavours, will find a final solution within the
Superstring Theory = Theory of Everything. In principle, it should allow us
to {\em calculate} all Yukawa couplings. Unfortunatelly, it is our lack of
understanding how the superstring can be linked in unambiguous way to lower
energy physics. Many theorists try to attack the problem in whole,
or its certain aspects, in the context of particular (among many billions)
superstring inspired models. However, the problem remains far away from
being solved and all what we know at present from superstring can be updated
in few important but rather general recommendations.

Nevertheless, one may rely that many aspects of the flavour problem can
be understood by means of more familiar symmetry properties. A relevant theory
could take place at some intermediate energies between the electroweak and
Planck scales. Nowadays the concepts of grand unification and supersymetry are
the most promising ideas towards the physics beyond the SM, providing a sound
basis for understanding the issues
of coupling unification and stability of gauge hierarchy. In particular, the
famous coupling crossing phenomenon in the Minimal Supersymmetric Standard
Model (MSSM)  points to the GUT scale $M_G\simeq 10^{16}$ GeV
\cite{Amaldi}. Applied to the flavour problem, these ideas should be
complemented by introducing some inter-family symmetry ${\cal H}$ that could
constrain the structure of fermion mass matrices. It is natural to expect
that ${\cal H}$ is also broken at the GUT scale $M_G$. Such a $SUSY
\otimes GUT \otimes {\cal H}$ theory can be regarded as a Grand Unification
of fermion masses. Here I shall review some aspects of the flavour
physics, and suggest a new possibility that could shed some more light
towards the search of such a theory.

The most difficult question is originated by the fact of family replication
itself. The measurement of the $Z$-boson decay width supports the idea
that there are just three observed families; at least, we are sure that
there are no more standard-like families with light neutrinos.
There seems to be no simple answer to the question "why three families?",
or "why {\em only} three families?". Apparently it is in the competence of
superstring, and I have nothing to add here. In what follows, I shall simply
accept that there are three families and pursue the
understanding of the issues (ii)-(v). Let us first outline these issues from
the SM point of view.

The mass spectrum of the quarks and charged leptons is spread over five orders
of magnitude, from MeVs to 100 GeVs \cite{Datagroup}.
In order to understand its shape it is
necessary to compare the fermion running masses at some scale $\mu\sim M_G$,
where the relevant new physics could take the place.\footnote{In what follows,
with an obvious notation we indicate by $u,d,...$ the fermion running masses
at the GUT scale, and by $m_u,m_d,...$ their physical masses. For the light
quarks (u,d,s) the latter traditionally are taken as running masses at $\mu=1$
GeV \cite{Leut}. } In doing so, we see that the {\em horizontal}
hierarchy of quark masses exhibits the approximate scaling low (see Fig. 1)
\be{qh}
t:c:u\sim 1:\eps_u:\eps_u^2\,,~~~~~~~
b:s:d\sim 1:\eps_d:\eps_d^2\,
\ee
where $\eps_u^{-1}=200-300$ and $\eps_d^{-1}=20-30$. As for the charged
leptons, they have a mixed behaviour:
\be{lh}
\tau :\mu : e\sim 1:\eps_d:\eps_u\eps_d
\ee
\vspace{6cm}
\begin{flushright}
\parbox{7cm}{{\sl Fig. 1. Logaritmic plot of fermion running masses at the
GUT scale versus family number. Points corresponding to the fermions with the
same electric charge are joined. The value $m_t$=130 GeV has been assumed. }}
\end{flushright}

\newpage

One can also observe that the {\em vertical} mass splitting is small within
the first family of quarks and is quickly growing with the family number:
\be{vert}
\frac{u}{d}\sim \frac{1}{2}\,,~~~~\frac{c}{s}\sim 8\,,~~~~
\frac{t}{b}\sim 60\,,
\ee
whereas the splitting between the charged leptons and down quarks
(at large $\mu$) remains considerably smaller (see Fig. 1). Moreover,
we observe that the third family is almost unsplit, $b\approx \tau$,
whereas the first two families are split but $d s\approx e \mu$.

One can also exploit experimental information on the quark mixing.
The weak transitions dominantly occur inside the families, and are suppressed
between different families by the small Cabibbo-Kobayashi-Maskawa
(CKM) angles \cite{Nir}:
\be{CKM}
s_{12}\sim \eps_d^{1/2}\,,~~~~ s_{23}\sim \eps_d\,,~~~~
s_{13}\sim \eps_d^{2}\,,
\ee
This shows that the quark mass spectrum and weak mixing pattern are strongly
correlated. Moreover, there are intriguing relations between
masses and mixing angles, such as the
well-known formula $\,s_{12}=\sqrt{d/s}\,$ for the Cabibbo angle.

All of the observed CP-violating phenomena \cite{Nir2} can be successfully
described in the frames of the SM due to sufficiently large ($\delta\sim 1$)
CP-phase in the CKM matrix.\footnote{It is tempting to mention that even an
outstanding issue of the cosmological baryon asymmetry can be accounted
entirely (with correct sign and magnitude) within the minimal SM, for the
reasonable values of the CKM angles and CP-phase \cite{Shaposh}. }
This means that the fermion mass matrices
are complex. The strong CP problem is closely related to this issue:
the net phase of the complex mass matrices should effectively contribute
to the P and CP violating $\Theta$-term, whereas
an absence of the neutron dipole electric moment puts
a strong bound $\Theta < 10^{-9}$ \cite{StrCP}.

As for neutrino masses and mixing, only some experimental upper bounds are
in our disposal \cite{Datagroup}. If SM is true all the way up to Planckian
energies, the neutrinos would stay massless, or could get some tiny masses
($\sim 10^{-5}$ eV) due to non-perturbative quantum gravitational (or
superstring) effects \cite{Planck}. However, some rather astrophysical
data, as are the solar and atmospheric neutrino deficites or the "after COBE"
evidence for some {\em hot} fraction of the cosmological dark matter,
may hint in favour of rather heavier and substantially mixed neutrinos.

As noted above, in the SM the fermion mass and mixing problem can be phrased
as a problem of the Yukawa coupling matrices: there is no explanation, what
is the origin of such a strong hierarchy of their eigenvalues, why they are
alligned approximately for the up and down quarks, what is the origin of their
complex structure, why the $\Theta$-term is vanishing in spite of this
complex structure etc. On the other hand, we believe that the SM
(or rather MSSM \cite{Kane}) is literally correct at lower energies.
This provides a
"boundary" condition for any hypothetical theory of the flavour: in the low
energy limit it should reduce to the minimal SM (or MSSM) in
{\em all} sectors, i.e. all the possible extra degrees of freedom
must decouple. Since the decoupling is expected to occur
at superhigh ($\sim M_G$) energies, there is practically no hope to observe
experimentally any direct dynamical effect of such a theory of the flavour.
It could manifest itself only in the sense of the flavour statics, through
certain constraints on the Yukawa sector of the resulting SM, or, in other
words, through the testable predictions for the fermion masses and CKM
parameters. In particular, it is tempting to think that the certain structure
of the mass matrices is responsible for several mass relations between
fermions and the CKM angles can be expressed as a functions of these
masses. It is also suggestive to think that these functions have the following
"analytic" properties \cite{decoupling}:

\underline{\em  Decoupling Hypothesis}. The mixing angles of the first
quark family with
others ($s_{12}$, $s_{13}$)  vanish in the limit $u,d\rightarrow 0$. At
the next step, when $c,s\rightarrow 0$, $s_{23}$ also vanishes.

\underline{\em Scaling Hypothesis}. In the limit when the masses of the
up and down quarks are proportional to each other: $u:c:t=d:s:b$, all mixing
angles ($s_{12}$, $s_{13}$ and $s_{23}$) are vanishing.

Therefore, our purpose is to find a self-consistent, complete and elegant
enough example of such a theory that could provide, besides solving other
fundamental problems, a natural explanation
to the fermion mass and mixing pattern.

\vspace{10mm}
\begin{flushright}
{\em "Models? No problem. We have many models."} \\
{\sl R.Mohapatra} \\
\end{flushright}
\vspace{-2mm}
{\bf 2. Mass Matrix Models}
\vspace{3mm}

Even the simplest extensions of the SM symmetry $G_{SM}=SU(3)_C\otimes SU(2)_L
\otimes U(1)_Y$ can provide some interesting hints for the understanding of
fermion mass and mixing pattern. For example, already the minimal grand
unified theory $SU(5)$ \cite{GUT} points to the possible origin of
$b\approx\tau$, specifying also that this relation should take place at the
GUT scale $M_G$. The SUSY $SU(5)$ theory, which is more appealing from the
viewpoint of internal consistency, leads to better quantitative agreement for
the $b-\tau$ unification,  as well as for the gauge coupling
unification \cite{Pokorski}. Unfortunately, the analogous relations
$s=\mu$ and $d=e$ are simply wrong, and one is forced to invoke some extra
sources (e.g. Planck scale induced higher dimensional
operators \cite{Ellis}), in order to split these masses from each other.
also, there is no hint neither for the origin of the fermion mass
hierarchy nor for the smallnes of the quark mixing angles: in the $SU(5)$
theory, as well as in the SM, the Yukawa couplings for the up and down type
fermions are different and there is no reason for their allignment.

Another minimal extension of the SM, so called $L\!-\!R$
model $G_{LR}=SU(2)_L\otimes SU(2)_R\otimes U(1)_{B-L}$
provides an "isotopic" symmetry interchanging the
up and down fermions, so that their mass matrices are somehow alligned.
Therefore, the smallness of quark mixing angles can be naturally linked
to the horizontal hierarchy of he quark masses, though the origin of the
hierarchy itself is beyond the scope of this model. Also, an additional
discrete $L\leftrightarrow R$ symmetry, essentially P-parity,
can be imposed naturally for the constraining of fermion mass matrices.

On the other hand, the $G_{LR}$ model does not imply any relations between
the quark and lepton masses, as the $SU(5)$ does. Its further extension to
the Pati-Salam theory $G_{PS}=SU(4)\otimes SU(2)_L\otimes SU(2)_R$
unifies the leptons with quarks as a fourth colour, and thereby provides
the possibility for the $b-\tau$ unification. However, it does not determine
the scale of this unification.

The ends are closed in the $SO(10)$ GUT (or rather SUSY GUT),
which accomodates each family of quarks and leptons of both chiralities
within the spinorial representation 16. $SO(10)$ is a logical final
towards both chains of the SM extensions:
$G_{SM}\rightarrow SU(5) \rightarrow SO(10)$, or $G_{SM}\rightarrow G_{LR}
\rightarrow G_{PS} \rightarrow SO(10)$. Therefore, the $SO(10)$ naturally
contains all types of the simplest fermionic symmetries: the isotopic and
quark-lepton symmetries as well as automatic P-parity.

All these appear to be necessary but not sufficient tools for the fermion
mass model building: also some inter-family (horizontal) symmetries
should be invoked in order to constrain the fermion mass matrices at the
needed degree. In the literature there are two main directions in the flavour
physics, which generally do not have strong intersection. These are: (i)
mass matrix ansatzes, and (ii) radiative mass generation.

The general aim of the first direction \cite{Weinberg,Fritzsch,Jarlskog} is
to provide {\em predictivity},
i.e. certain relations between fermion masses and the CKM angles, by
constraining the mass matrix form and by reducing the  number of its
parameters. This can be motivated
by some family symmetries (or, in some cases, even are not motivated).
In general, this implies a study of the so called "zero textures" - matrices
with the certain number of zero elements. In order to reduce a number of
arbitrary parameters, together with horizontal symmetry one should utilize
the above mentioned GUT ingredients as are the isotopic and quark-lepton
symmetry and P-parity.
One of the most interesting mass matrix ansatzes is given by Fritzsch texture
\cite{Fritzsch}
\be{Fr}
\hat{m}_f=\,\matr{0}{A_f}{0}{A_f'}{0}{B_f}{0}{B'_f}{C_f}\,,~~~~~~f=u,d,e
\ee
which can be obtained at the price of some horizontal symmetry. This structure
implies that the fermion mass generation starts from the heavist $3^{rd}$
family ($C$ is a largest entry in eq. (\ref{Fr})) and proceeds to lighter
families through the mixing terms. In the context of the $L\!-\!R$ symmetric
model this matrices can be Hermitian due to P-parity.
Thereby, neglecting the phase factors, the total number of the parameters for
3 mass matrices ($f=u,d,e$) is reduced to a number 9, i.e. just
to the number of quarks and leptons. This allows to express the quark mixing
angles in terms of their mass ratios: in particular, for the Cabibbo angle
we can obtain $s_{12}=\sqrt{d/s}$. Unfortunately, the Fritzsch texture
seems to be already excluded by recent CDF bound on the top mass.
However, there are some other suggestions \cite{Jarlskog} which still agree
to the experimental data.

The idea of radiative mass generation \cite{rad1} in general aims to
provide rather quolitative explanation to the fermion mass hierarchy.
Indeed, it is tempting to think that the $1\!-\!2$ orders of magnitude
hierarchy between fermion masses (see eqs. (\ref{qh}) and (\ref{lh})) is
due to loop expansion: $\eps\sim (h^2/16\pi^2)$ with $h$ being a typical
Yukawa coupling of the order of 1. This could be, if due to some symmetry
reasons only the heaviest $3^{rd}$ family gets mass at tree level, whereas
the $2^{nd}$ becomes massive at the 1-loop level and the $1^{st}$ only at
2-loops. The models exhibiting this feature, were suggested in
\cite{Balakr,Rat}. However, the radiative models fail in predictivity.
Moreover, it is rather difficult to obtain a quantitatively consistent
picture, and also to avoide dangerous flavour changing phenomena
\cite{Rat,NOP,Rat2}.

The general feature of the frameworks considered above is that the
mass generation starts from the heaviest third family and then propagates
to the lighter ones. However, the fermion mass pattern may hint that the
case is just the opposite, and the first family plays an unique role in
mass generation.
We stick to the observation that the GUT scale running masses of the
electron, u-quark and d-quark are not strongly split, which maybe manifests
the approximate symmetry limit. With this picture in mind, it is suggestive
to think that the masses of the $1^{st}$ family are somehow related to an
energy scale $M_1$ at which this symmetry is still good, while the masses
of the $2^{nd}$ and $3^{rd}$ family are respectively related to lower scales
$M_2$ and $M_3$ at which it is no longer as good. Suppose that the first
family is indeed the starting point, and that the expression like eq.
(\ref{qh}) holds for the inverse masses rather then masses, namely
\be{inverse}
\frac{1}{f_i}\sim \frac{\eps_f^{i-1}}{m}\,,~~~~~~f=u,d,e
\ee
where $i=1,2,3$ is a family number. Then we have $u\!\sim\! d\!\sim \!m$,
$c/s\!\sim (\eps_d/\eps_u)>1$ and $t/b\sim (\eps_d/\eps_u)^2\gg 1$. In this
way, the splitting between up and down quark masses in Fig. 1 is understood
by means of one parameter $\eps_d/\eps_u>1$. We call the above formula the
{\em inverse hierarchy pattern}.

The above consideration suggests that the mass generation proceeds from
the lightest family to heavier ones.
At first sight, it is nonsense. However, we do not imply this literally.
Let us take here some break and go back to neutrinos.

The $SO(10)$ extension (in fact, already the $L\!-\!R$ symmetric model)
brings new particles in addition
to the minimal fermion spectrum of the SM. These are right handed neutrinos.
However, it is not expected that they will be seen at lower energies.
After $SO(10)$ breaking down to SM no symmetry defences them to be massless,
so they should acquire $O(M_G)$ Majorana mass terms and thereby decouple from
the light particle spectrum. On the other hand, the
$SO(10)$ invariant Yukawa couplings provide the neutrino Dirac mass terms,
which in the standard picture resemble the up quark masses. The
interplay of both mass structures results in the famous seesaw mechanism,
which provides the naturally small majorana masses to physical neutrinos
\cite{seesaw}. The resulting neutrino mass matrix reads as
\be{seesaw}
\hat{m}_{\nu}=v^2 \Gamma \hat{M}_R^{-1} \Gamma^T
\ee
where $\hat{M}_R$ is a Majorana mass matrix of right handed neutrinos and
$\Gamma$ is a matrix of the "Dirac" Yukawa couplings.

One can imagine, that there are also some charged fermion states which
are allowed to be superheavy by GUT symmetry. For example, already further
extension of the $SO(10)$ to the $E_6$ model brings such heavy states.
Then it is suggestive to think that the masses of ordinary quarks and
leptons could appear through the ana\-logous seesaw mixing with these
heavy states. Such a possibility, named subsequently as
an {\em universal seesaw mechanism}, was suggested in
\cite{uniseesaw}. Then for the quark and lepton mass matrices we have
the expression analogous to eq. (\ref{seesaw}).

It seems quite natural to assume that the Yukawa couplings all are $O(1)$
and the
fermion mass hierarchy is initiated in the heavy fermion sector, while the
usual light fermions are just the spectators of this phenomenon. Then
the inverse power in eq. (\ref{seesaw}) is crucial: by means of the seesaw
mechanism this hierarchy will be transfered to ordinary fermions
in an inverted way.\footnote{The idea of universal seesaw mechanism was also
explored in a number of papers \cite{uniseesaw2}. The inverse hierarchy,
however, corresponds to the spirit of the original paper \cite{uniseesaw},
where it was in fact first suggested. }
This can provide a firm basis to the
inverse hierarchy pattern of the eq. (\ref{inverse}). It is clear,
that the heaviest ones among the heavy fermions are just the partners of the
$1^{st}$ standard family, and its small mass splitting can be just a
reflection of the symmetry limit: namely, these heaviest of heavies can be
so heavy \cite{soheavy}, in particular, heavier than the relevant GUT scale,
that their mass terms obey the isotopic and quark-lepton symmetries,
which are the natural subsymmetries e.g. of the $SO(10)$.

The seesaw induced inverse hierarchy pattern was explored in radiative mass
generation scenario for quarks \cite{Rat2,rattazzi2} and also included
leptons \cite{rattazzi3}. As a result, several intriguing predictions were
obtained for the fermion masses and mixing angles. It is clear, however,
that the use of a radiative mechanism to generate the mass hierarchy in a
heavy fermion sector is in obvious contradiction with the idea of low-energy
supersymmetry. Within SUSY scheme one should think of some tree level
mechanism that could generate the masses of heavy fermions by means of the
effective operators of successively higher dimension, thus providing a
hierarchical structure to their mass matrix.

Before proceeding let us comment also, that universal seesaw can
automatically solve the strong CP-problem without
introducing an axion, {\em a l\'a} Nelson-Barr mechanism \cite{Nelson}.
Such models were suggested in \cite{BM,Berez} on the basis of the
spontaneously broken P-parity \cite{BM} or CP-invariance \cite{Berez},
where the $\Theta$-term is automatically vanishing at tree level and
appears to be naturally small due to loop corrections. Alternatively,
within the seesaw picture one can naturally incorporate the Peccei-Quinn
type symmetries \cite{uniseesaw,Khlopov}, where the axion appears to be
simultaneously a majoron.

\newpage


{\bf 3. Inverse Hierarchy in SUSY ${\bf SO(10)}$ Model}
\vspace{3mm}

We intend to built a predictive SUSY $SO(10)$ model for the fermion masses,
pursuing the universal seesaw mechanism in order to obtain naturally the
inverse hierarchy pattern.
For this purpose one has to appeal to some symmetry properties.
We suggest that there is some "family-type" symmetry (discrete or global)
${\cal H}$, that distinguishes the superfields involved into the game.
In the following we will not specify the exact form of ${\cal H}$, describing
only the pattern how it should work.
We also wish that our model fulfills the following fundamental conditions:

{\em A. Unification of the strong, weak and weak hypercharge gauge couplings}
$-$ correct prediction for $\sin^2\!\theta_W$ or $\alpha_s$ at lower energies.

{\em B. Natural (not fine-tuned) gauge hierarchy and doublet-triplet
splitting} $-$ a couple of Higgs doublets should remain light whereas
their colour triplet partners in GUT supermultiplet must be superheavy.

{\em C. Sufficiently long-lived proton} $-$ proton lifetime should be above
the recent experimental lower bound $\tau_p>10^{32}$ yr.

{\em D. Natural suppression of the} FCNC.

Let us design such a SUSY $SO(10)\otimes {\cal H}$ model.
We know that three families of quarks and leptons should be arranged within
16-plets $16^f_i$, $i=1,2,3$. Besides them, I exploit three families of
superheavy fermions $16^F_k+\overline{16}^F_k$. All these have certain
transformation properties under ${\cal H}$ symmetry. For the following it is
convenient to describe them in the terms of $SU(4)\otimes SU(2)_L\otimes
SU(2)_R$ decomposition:
\be{16f}
16^f_i=f_i(4,2,1)+f^c_i(\bar{4},1,2)
\ee
\be{16F}
16^F_i={\cal F}_i(4,2,1)+F^c_i(\bar{4},1,2) \,,~~~~~~~~
\overline{16}^F_i={\cal F}^c_i(\bar{4},2,1)+F_i(4,1,2)
\ee
For the electroweak symmetry breaking and quark and lepton mass generation
we use a traditional 10-dimensional Higgs supermultiplet
\be{10}
10=\phi(1,2,2)+T(6,1,1)
\ee
For the $SO(10)$ symmetry breaking we promote, as usual,
a set of scalar superfields, consisting of various 54-plets,
45-plets and 126-plets, which also have different transformation
properties under ${\cal H}$.
\beqn{scalars}
54=(1,1,1)+(1,3,3)+(20,1,1)+(6,2,2)~~~  \nonumber \\
45=(15,1,1)+(1,3,1)+(1,1,3,)+(2,2,6)~~ \\
126=(10,1,3)+(\overline{10},3,1)+(6,1,1)+(15,2,2) \nonumber
\eeqn
We suggest that all 54-plets have the VEVs of standard configuration
corresponding to the symmetry breaking channel $SO(10)\rightarrow
SU(4)\otimes SU(2)_L\otimes SU(2)_R$. As for the 45-plets, we suggest that
there are three types of them: $45_{BL}$-type fields with VEV on their
(15,1,1) fragment, breaking $SU(4)$ down to $SU(3)_c\otimes
U(1)_{B-L}$; $45_R$-type fields with VEV on the (1,1,3) component, providing
the breaking $SU(2)_R\rightarrow U(1)_R$; and $45_X$-type fields having the
VEVs on both (15,1,1) and (1,1,3) components:
\be{vevs}
\begin{array}{l}
\langle 54 \rangle = I\otimes \mbox{diag}(1,1,1,-3/2,-3/2)\cdot V_G \\
\langle 45_{BL} \rangle = \sigma\otimes \mbox{diag}(1,1,1,0,0)\cdot V_{BL} \\
\langle 45_R \rangle = \sigma\otimes \mbox{diag}(0,0,0,1,1)\cdot V_R~~; \\
\langle 45_X \rangle = \sigma\otimes \mbox{diag}(1,1,1,x,x)\cdot V_X
\end{array} ~~~~
I\!=\!\mat{1}{0}{0}{1}\!,~~~\sigma\!=\!\mat{0}{1}{-1}{0}
\ee
Finally, the 126-plet with VEV $v_R$ across the (10,1,3) direction completes
the $SO(10)$ breaking down to $G_{SM}=SU(3)_c\otimes SU(2)_L
\otimes U(1)_Y$.
Motivated by the coupling crossing phenomenon in MSSM \cite{Amaldi}, we
suggest that $SO(10)\otimes {\cal H}$ breaks down to $G_{SM}$ at once,
by VEVs $V_G,V_X,V_{BL},v_R\sim M_G$. Below this scale the
theory is just MSSM, with three fermion families ($f_i$) and
one light couple of the Higgs doublets ($\phi$).
Many questions of the series {\em A - D} are immediately respected in this
way: $\sin^2\!\theta_W(\mu)$ and $\alpha_s(\mu)$ are
correctly related at $\mu=M_Z$; the FCNC are naturally suppressed
provided that SUSY breaking sector has simple (e.g. universal) structure;
large unification scale ($M_G\simeq 10^{16}$ GeV) saves the proton from
the unacceptably fast decay mediated by $X,Y$ gauge bosons ($d=6$ operators).
Let us comment also, that the SUSY $SO(10)$ theory has automatic matter
parity, under which the spinorial representations change the sign while the
vectorial ones stay invariant.\footnote{In fact, this gives a natural ground
to refer the spinorial representations as fermionic superfields, and the
vectorial ones as Higgs superfields.}
Provided that non of the $16$-plets have the
VEV, this implies an automatic $R$-parity conservation for the resulting
MSSM. It is well-known, that proton decaying $d=4$ operators
mediated by squarks are vanishing in this case.

In order to establish the seesaw regime at once, we assume that
${\cal H}$ symmetry does not allow $16^f16^f 10$ couplings but only the
following terms in the Yukawa superpotential:
\be{WY}
\Gamma_{ik} 10\, 16^f_i 16^F_k + G_{ik} 45_R\, 16^f_i \overline{16}^F_k
\ee
These couplings generate the mass terms for ordinary quarks and leptons
($f$-fermions) by means of seesaw mixing with the
$F$-fermion states afterthat the latter
become superheavy. The whole $9\times 9$ mass matrices have the form
\be{Mtot}
\begin{array}{ccc}
 & {\begin{array}{ccc} \,f^c\,\, & \,\,\,\;F^c & \,\,\,\,\;{\cal F}^c
\end{array}}\\ \vspace{2mm}
M^f_{tot}~=~\begin{array}{c}
f \\ F \\ {\cal F}   \end{array}\!\!\!\!\!&{\left(\begin{array}{ccc}
0 & \hat{M}_{fL} & 0 \\ \hat{M}_{fR}  & \hat{M}_F & 0 \\
\hat{M}^{\dagger}_{fL} & 0 & \hat{M}_{\cal F} \end{array}\right)}
\end{array}
\ee
Where $\hat{M}_{fL}=\Gamma\langle\phi\rangle$ and
$\hat{M}_{fR}=G^T \langle 45_R\rangle$.
We do not specify further the form of the Yukawa couplings.
We only suggest that they all are $O(1)$, as well as the gauge coupling
constants. $\Gamma$ and $G$ are some general complex and non-degenerated
matrices. Without loss of generality, by suitable simultaneous redefinition
of the basis of $f$ fermions of all types ($f=u,d,e,\nu$), we always can
bring them to a skew-diagonal form:
\be{skew}
\Gamma_{ik},\,G_{ik}=0\,,~~~~~    \mbox{if}~~~ i<k.
\ee
The role of the Higgs 10-plet is crucial. We require that its
$\phi(1,2,2)$ component, which consists of the Higgs doublets, remains
massless in the SUSY limit. On the other hand, the $T(6,1,1)$ fragment,
containing colour triplets, should acquire the mass of the order of $M_G$:
otherwise it would cause unacceptably fast proton decay and would affect
the unification of the gauge couplings. In order to resolve this famous
problem of the doublet-triplet splitting without fine tuning of the
superpotential parameters, one can address to the
Dimopoulos-Wilczek mechanism, utilizing the Higgs $45_{BL}$-plet \cite{DiWi}.
The VEV of $\phi$ arises after the SUSY breaking and
breaks the $SU(2)_L\otimes U(1)_Y$ symmetry:
\be{phi}
\langle\phi\rangle=\mat{v_2}{0}{0}{v_1},~~~~~~~(v_1^2+v_2^2)^{1/2}=v=
175\,\mbox{GeV}
\ee
This implies that the $(1,2)$-blocks of the matrix $M^f_{tot}$ are
essentially the same: $\hat{M}_{fL}=\Gamma v\sin\!\beta\,$ for the up-type
fermions ($f=u,\nu$) and $\hat{M}_{fL}=\Gamma v\cos\!\beta$ for the down-type
ones ($f=d,e$), where $\tan\!\beta=v_2/v_1$ is the  famous up-down VEV ratio
in MSSM.

Equally important is the choice of the VEV $\langle 45_R\rangle$ towards
the $(1,1,3)$ direction.
It tells that the $(2,1)$-block of $M^f_{tot}$ differs {\em only}  by the
sign for the up-type and down-type fermions: $\hat{M}_{fR}=+G^TV_R$ for
$f=u,\nu$ and $\hat{M}_{fR}=-G^TV_R$ for $f=d,e$. On the other hand,
it implies that the $(1,3)$-block is vanishing. Therefore, only
the $SU(2)_L$-singlet $F$-type fragments of eq. (\ref{16F}) are important
for the seesaw mass generation, whereas the ${\cal F}$-type ones are
irrelevant. As it was shown in \cite{proton}, this feature is decisive for
the natural suppression of the dangerous $d=5$ operators for the proton
decay.\footnote{Indeed, the $f$ and ${\cal F}$ states are unmixed, so
the colour triplets in $T(6,1,1)$ can cause transitions of $f$'s
only into the superheavy ${\cal F}$'s. Therefore, the baryon number violating
$d=5$ operators $[ffff]_F$ (so called $LLLL$ type operators), which bring
the dominant contribution to the proton decay after dressing by the
$\tilde{W}$-bosinos, are automatically vanishing. As for the $RRRR$ type
operators $[f^cf^cf^cf^c]_F$, they clearly appear due to
the $f^c-F^c$ mixing. But they are known to be much more safe for the proton
(see e.g. \cite{Nath} and refs. therein). }

Therefore, we have all the key ingredients for the inverse hierarchy
ansatz. What remains is to obtain the needed hierarchical pattern
for the heavy mass matrices $\hat{M}_F$. Let us assume that the
${\cal H}$ symmetry allows the bare mass term ($M\gg M_G$) for the $1^{st}$
heavy family $F_1$ and the mass of the $2^{nd}$ one ($F_2$) is generated via
$45_X$:
\be{F12}
M16^F_1\overline{16}^F_1\,+\,g45_X16^F_2\overline{16}^F_2 \,,
\ee
while the $3^{rd}$ family becomes massive through the effective operator
$(45^2_X\!/M\!)16^F_3\overline{16}^F_3$.
In this case the fermion mass hierarchy will be explained due to small
parameter $\eps\!\sim \!V_G/M$. However, it is not enough restrictive to use
this later operator without defining to which of the possible $SO(10)$
channels it acts: $45\times 45\rightarrow 1+45+210$.
In order to be less vague, let us introduce the
additional couple $16^F_0+\overline{16}^F_0$ with barr mass
$M'\!\sim\! M$ and Yukawa couplings $g'16^F_3\overline{16}^F_0 +
g''16^F_0\overline{16}^F_3$.
\footnote{For the simplicity we assume that $F_0$ has
no couplings with the $16^f$'s, though it is easy to see that such couplings
would not affect significantly our results. The possible contributions
of the 10-plet couplings to heavy states (\ref{16F}) are also negligible. }
Then the mass terms of the $F_3$ states appear at the decoupling of the
heavier $F_0$ states, i.e. after the diagonalization of the mass matrix
\be{MF3}
\begin{array}{cc}
 & {\begin{array}{cc} F^c_3\;\;\; &\;\;\; F^c_0 \end{array}}\\
\begin{array}{c} F_3 \\ F_0 \end{array}\!\!\!\!\! &{\left(\begin{array}{cc}
0 & g'\langle 45_X\rangle \\ g''\langle 45_X\rangle & M'\end{array}
\right)}\end{array}
\ee
As a consequence, we obtain the mass matrices $\hat{M}_F$ of the desired form:
\be{MassF}
\hat{M}_F=M(\hat{P}_1+\eps_f\hat{P}_2+\eps_f^2\hat{P}_3)\,,
{}~~~~~~~~~\begin{array}{c} \hat{P}_1=\mbox{diag}(1,0,0) \\
\hat{P}_2=\mbox{diag}(0,1,0) \\ \hat{P}_3=\mbox{diag}(0,0,C) \end{array}
\ee
where $C\!\sim \!M/M'\!\sim \!1$, since all Yukawa couplings are assumed
to be $O(1)$. What is new, is that the complex
expansion parameters $\eps_f$ ($f=u,d,e,\nu$) {\em are not} independent
anymore, but are related due to the VEV pattern (\ref{vevs}) of the $45_X$:
\beqn{eps4}
\eps_d=\eps_1+\eps_2\,,~~~~~~\eps_e=-3\eps_1+\eps_2  \nonumber \\
\eps_u=\eps_1-\eps_2\,,~~~~~~\eps_\nu=-3\eps_1-\eps_2
\eeqn
from where follows
\be{eps2}
\eps_e=-\eps_d-2\eps_u\,,~~~~~~~~ \eps_\nu=2\eps_e+3\eps_u
\ee
Assuming $\hat{M}_F\gg GV_R$, the seesaw block-diagonalization of eq.
(\ref{Mtot}) results in following mass matrices for the ordinary quarks
and leptons:
\be{mf}
\hat{m}_f=\zeta_fvV_R\Gamma\hat{M}_F^{-1}G^T
\ee
where
$\zeta_f=\sin\!\beta\,$ for $f=u,\nu\,$ and $\zeta_f=-\cos\!\beta\,$ for
$f=d,e$. In this way the inverse proportionality of eq. (\ref{inverse}) is
realized.
The seesaw limit $M_F\gg V_R$ is certainly very good for all light states
apart from $t$-quark,
since their masses must be much smaller than $v$. However, since
$m_t=O(v)$, we expect the mass of its F-partner
$M_T$ to be of the order of $V_R$ (remember that the Yukawa couplings are
considered to be $O(1)$).\footnote{Decoupling of the heavy states $F$ occurs
at the scale $V_R$: below this scale the effective theory is the MSSM, and
$V_R\Gamma\hat{M}_F^{-1}G^T=\hat{m}_f/v\zeta_f$ are in fact the MSSM Yukawa
couplings. Then the ratio $V_R/M$ is given by $m_u/v\sim\!10^{-5}$. Taking
into account that $M/M_G\sim \eps_d\sim 30$, this implies $V_R\sim 10^{13}$
GeV, i.e. some three order of magnitude below the GUT scale $M_G$. This is
not a big problem neither for gauge coupling unification nor for other issues,
provided that the VEV of the $126$-plet $v_R$ is of the order of $M_G$.
Nevertheless, one may does not consider such a small $V_R$ as enough appealing.
In this case we can suggest that the $(2,1)$-block $\hat{M}_{fR}$ of the
"big" mass matrix (\ref{Mtot}) appears due to the effective operators
$(45_R^2/M) 16_f \overline{16}_F$ rather than the direct Yukawa couplings of
the eq. (\ref{WY}). These operators can be built in the same manner as we did
for the third heavy family $F_3$. Then $\hat{M}_{fR}\sim 10^{13}$ GeV can
occur naturally for $V_R\sim M_G$, and non of our results will change. }
Thus, to evaluate $m_t$, we need the mass matrix
without the  restriction $\hat{M}_F\gg V_R$. This is given by
\be{exactseesaw}
\hat{m}_f \hat{m}_f^{\dagger} = (\zeta_fv)^2 \Gamma
\left[1\,+\,\hat{M}_F^{\dagger}(G^TG^{\ast}V_R^2)^{-1} \hat{M}_F \right]^{-1}
\Gamma^{\dagger} \,.
\ee
Notice that, when $V_R\gg\hat{M}_F$, this equation gives the obvious
result $\hat{m}_f=\Gamma \zeta_f v $. On the other hand, when
$V_R\ll \hat{M}_F$, it reduces to the seesaw formula (\ref{mf}).

It is useful to first study $\hat{m}_f$ in the seesaw limit (\ref{mf}).
The exact formula (\ref{exactseesaw}) will only be relevant to evaluate $m_t$.
Once again, the inverse matrices are easier to analyse.
{}From the eqs. (\ref{MassF}) and (\ref{mf}) we have
\be{mf-1}
\hat{m}_{f}^{-1}=\frac{M}{v\zeta_fV_R}(G^T)^{-1}
(\hat{P}_1+\eps_f\hat{P}_2+\eps_f^2\hat{P}_3)\Gamma^{-1}=
\frac{1}{m\zeta_f}(\hat{Q}_1+\eps_f\hat{Q}_2+\eps_f^2\hat{Q}_3)\,,
\ee
where
$\hat{Q}_n\propto (G^T)^{-1}\hat{P}_n\Gamma^{-1}$ are still rank-1
matrices, but not orthogonal anymore. We can also choose a basis
of eq. (\ref{skew}) and use a relation
$(G^T)^{-1}\hat{P}_1\Gamma^{-1}=(G_{11}\Gamma_{11})^{-1}\hat{P}_1$,
to define $\hat{Q}_1=\hat{P}_1$ and
$m=\Gamma_{11}G_{11} v V_R/M$. In other words, without loss of generality
we can take
\be{Q}
\hat{Q}_1\!=\!(1,0,0)^T\!\!\bullet(1,0,0), ~~~
\hat{Q}_2\!=\!(a,b,0)^T\!\!\bullet(a'\!,b'\!,0), ~~~
\hat{Q}_3\!=\!(x,y,z)^T\!\!\bullet(x'\!,y'\!,z')
\ee
so that the inverse mass matrices at the leading order are the following:
\be{form}
\hat{m}_f^{-1}\,=\frac{1}{m\zeta_f}\matr{1+aa'\eps_f}{ab'\eps_f}{xz'\eps_f^2}
{ba'\eps_f}{bb'\eps_f}{yz'\eps_f^2}{zx'\eps_f^2}{zy'\eps_f^2}
{zz'\eps_f^2}
\ee
Here have we neglected $O(\eps)$ order corrections in every element except
the $11$-one. In order to split fermion masses within the
first family and accomodate large ($\sim\!\sqrt{\eps_d}$) Cabibbo angle,
the matrix (\ref{form}) must be diagonalized considering that $aa'\eps_{d,e}\!
\sim\!1$.\footnote{One may wonder how to achieve $\eps_daa'\simeq 1$, if the
Yukawa couplings are assumed to be $O(1)$
and $\eps$ is a small parameter: $\eps_d\!\sim\!1/20\!-\!1/30\,$ (see eq.
(\ref{qh})). However, here we still see the advantage of seesaw mechanism:
$a$ and $a'$ are not the coupling constants but rather their ratios,
due to the "sandwiching" between $\Gamma$ and $G$ in eq. (\ref{mf}).
Thus, it should not come as a surprise if $aa'\!\sim\! 20-30$ due to some
spread in the Yukawa coupling constants (for example, if both
$a=\Gamma_{21}/\Gamma_{11}$ and $a'\!= G_{21}/G_{11}$ are $\sim\!4\!-\!5$),
while the Yukawa constants themselves are small enough to fulfill the
triviality bound $G^2_{Y}/4\pi<1$. On the other hand, the pattern
of the fermion masses and mixing suggests that such a "coherent" enhancement
does not happen for other entries in the matrix (\ref{mf}), so that the
corresponding $O(\eps)$ corrections are negligible.  }

Thus, for the fermion mass eigenvalues at the GUT scale we have
\beqn{eigen}
\frac{m\sin\!\beta}{u}=|1+\eps_u aa'|\,,~~~~~
\frac{m\sin\!\beta}{c}=\frac{|\eps_u bb'|}{|1+\eps_u aa'|}\,,~~~~~
\frac{m\sin\!\beta}{t}=|\eps_u^2 zz'| \nonumber \\
\frac{m\cos\!\beta}{d}=|1+\eps_d aa'|\,,~~~~~
\frac{m\cos\!\beta}{s}=\frac{|\eps_d bb'|}{|1+\eps_d aa'|}\,,~~~~~
\frac{m\cos\!\beta}{b}=|\eps_d^2 zz'| \\
\frac{m\cos\!\beta}{e}=|1+\eps_e aa'|\,,~~~~~
\frac{m\cos\!\beta}{\mu}=\frac{|\eps_e bb'|}{|1+\eps_e aa'|}\,,~~~~~
\frac{m\cos\!\beta}{\tau}=|\eps_e^2 zz'| \nonumber
\eeqn
{}From the two first rows of eqs. (\ref{eigen}) we have
\be{top}
\left|\frac{\eps_u}{\eps_d}\right|=\frac{ds}{uc}\tan^2\!\!\beta\,,~~~
\left|\frac{\eps_u}{\eps_d}\right|^2=\frac{b}{t}\tan\!\beta ~~~
\Longrightarrow ~~~
\frac{t}{b}=\left(\frac{uc}{ds}\right)^2 \tan\!^{-3}\!\beta
\ee
This expression for the top mass is valid in the seesaw limit of eq.
(\ref{mf}). However, due to seesaw corrections,
it actually gives only an upper bound. Indeed,  by using the correct
mass matrix of eq. (\ref{exactseesaw}), the eq. (\ref{top}) is reduced to
\be{topineq}
t=\frac{bR}{\sqrt{1+(bR/\Gamma_{33}v\sin\!\beta)^2}}<bR\,;~~~~
R=(uc/ds)^2\tan\!^{-3}\!\beta
\ee
where for the perturbativity one can assume $\Gamma_{33}\leq 2$.
This equation is valid at the GUT scale and to discuss its
implications, the running of masses needs to be considered.
In doing so, it appears to be rather restrictive.
In particular, by taking $m_u/m_d\leq 0.7$ and $m_c/m_s\leq 12$ as upper
bounds and bearing in mind that $\tan\!\beta\geq 1$, we get
\be{Rmax}
R\leq R_{\sf max}\simeq 70
\ee
which sets an upper bound on the top physical mass $m_t$ of about 150 GeV.
On the other hand, by taking the recent CDF bound $m_t>109$ GeV we have
\be{Rmin}
R\geq R_{\sf min}=36
\ee
which translates into the strong upper bound $\tan\!\beta<1.1$
for the same values of $m_u/m_d$ and $m_c/m_s$. Alternatively, by assuming
$\tan\!\beta=1$, we have a lower bound  $uc/ds>6$. Then, by using
$m_c/m_s<12$, we get $m_u/m_d>0.5$.

{}From the second two rows of eqs. (\ref{eigen}) we can derive
the mass formulae
\be{bottom}
\sqrt{\frac{b}{\tau}}=\frac{ds}{e\mu}=\left|\frac{\eps_e}{\eps_d}\right|=
\left|1+2\,\frac{\eps_u}{\eps_d}\right|
\ee
When I get to the bottom I go back to the top \cite{Beatles}: the eq.
(\ref{top}) shows that the $\eps_u/\eps_d$ ratio is small:
$|\eps_u/\eps_d|=0.12\div 0.16$. Then the eq. (\ref{bottom}) approximately
gives the GUT relationships between the down quark and lepton masses:
\be{bds}
b=\tau\,,~~~~~~ds=e\mu
\ee
where $O(\eps_u/\eps_d)$ corrections bring about $30\,\%$ uncertainty.
Running down these relations from the GUT scale we get appealing
values for the down quark masses: $m_b=4-5$ GeV, $m_s=100-150$ MeV and
$m_d=5-7$ MeV, where for deriving the light quark masses we have used the
current algebra relation $m_s/m_d=20$. This later relation, however,
cannot be derived from our consideration itself. Moreover, it is not clear
whether it is consistent in our scheme: at first sight the relation
$|\eps_d|\approx |\eps_e|$ implies that $d/s\approx e/\mu\sim |\eps_e|$,
whereas the experimental values $s/d\approx 20$ and $\mu/e\approx 200$
differ by the order of magnitude.

The answer is "Yes"! Indeed, by taking into account that $\tan\!\beta\!
\approx\! 1$, we get:
\be{splitting}
u/d=|1+\eps_daa'|\,,~~~~~~
u/e=|1+\eps_eaa'|
\ee
and
\be{d/s}
\frac{(s/d)}{(\mu/e)}=\frac{|1+\eps_daa'|^2}{|1+\eps_eaa'|^2}
=\left(\frac{e}{d}\right)^2 =\left(\frac{s}{\mu}\right)^2
\ee
where we have neglected $O(\eps_u/\eps_d)$ corrections. Therefore, in
order to split the electron and $d$-quark masses from the $u$-quark mass
respectively by factors of about $\frac{1}{2}$ and 2 (see Fig. 1), we have
to assume that $|\eps_eaa'|\!\approx\! 1$.
The relation $\eps_d\!\approx\! -\eps_e$ is crucial, since it splits $d$ and
$e$ to different sides from $u\!\approx\!m$ by about a factor 2.
Then, according to eq. (\ref{d/s}), the order of
magnitude difference between $\mu/e$ and $s/d$ follows automatically. In this
way, owing to the numerical coincidence $(e/d)^2\!\sim\! \eps_u\eps_d$, we
reproduce the mixed behaviour of leptons (see eq. (\ref{lh})).
This is not, however, the end of the story. By assuming that $|\eps_eaa'|\leq
1$ (which, as we show below, can be derived by considering the quark mixing),
the eqs. (\ref{splitting}) and (\ref{d/s}) imply
\be{u/d}
\frac{m_u}{m_d}=|1+\eps_eaa'|\left(\frac{m_em_s}{m_\mu m_d}\right)^{1/2}
\leq 0.65
\ee
where the $O(\eps_u/\eps_d)$ corrections can cause about $20\,\%$ uncertainty
in this estimate.

Let us discuss now the pattern of the weak mixing. It is easy to see that
the quark mixing arises dominantly due to diagonalization of the down quark
mass matrix $\hat{m}_d$. The up quark matrix $\hat{m}_u$ is much more
"stretched" and essentially close to its diagonal form, so that it brings only
$O(\eps_u/\eps_d)$ corrections to the CKM mixing angles.
Let us denote by $\theta^{L,R}_{12}$, $\theta^{L,R}_{23}$ and
$\theta^{L,R}_{23}$ the angular parameters of the unitary matrices $V_{L,R}$
diagonalizing $\hat{m}_d$.
Then from the eqs. (\ref{form}) and (\ref{eigen}) we see that
\be{s12}
s^L_{12}=\frac{|\eps_d ab'|}{|1+\eps_d aa'|}\,, ~~~
s^R_{12}=\frac{|\eps_d a'b|}{|1+\eps_d aa'|}\, ~~\Longrightarrow ~~
s^L_{12}s^R_{12}=\frac{d}{s}\,|\eps_d aa'|
\ee
where "$s^{L,R}$" stand for $\sin\!\theta^{L,R}$.
Then, by assuming that the right-handed current "Cabibbo" angle  $s^R_{12}$
is not anomalously large (not larger than the ordinary Cabibbo angle
$s^L_{12}\!=\!s_{12}\!\approx\sqrt{d/s}$, which in itself is already much
larger than it was expected naively ($\sim \eps_d$) from the eq. (\ref{s12})),
we obtain that $|\eps_d aa'|\leq 1$.
On the other hand, we know that $|\eps_d aa'|$ should be
rather close to 1, in order to achieve a sufficient $d-u-e$ splitting.
By assuming that $s^R_{12}\sim s_{12}$,
the eq. (\ref{s12}) translates into
\be{Cabibbo1}
s_{12}\approx \sqrt{\frac{d}{s}\left|1-\frac{u}{d}e^{i\delta_d}\right|}\,,~~~~
{}~~\delta_d=\arg |1+\eps_d aa'|
\ee
Thus, the Cabibbo angle has to be in the right range:
$s_{12}\sim \eps_d^{1/2}$. Obviously, the values of other mixing angles also
fit parametrically the pattern of the eq. (\ref{CKM}): $s_{23}
\sim \eps_d$ and $s_{13}\sim \eps_d^{2}$. Without exploiting the
concrete structures of the Yukawa coupling matrices $\Gamma$
and $G$ it is not possible to make exact predictions for the CKM matrix
parameters. It is natural to expect, however, that the ${\cal H}$ symmetry
will constrain somehow the form of $\Gamma$ and $G$, and thereby will enhance
the predictivity. Let us assume, for example, that $\Gamma$ has
the Fritzsch form (\ref{Fr}), involving 5 parameters. Then, rotating the
fields to the skew-diagonal basis of eq. (\ref{skew}), we see that 6 complex
entries of $\Gamma$ are not independent anymore, but are related through
$\Gamma_{31}=-\Gamma_{21}\Gamma_{22}\Gamma_{32}^{-1}$. The same can occur
for the matrix $G$. Then we immediately receive an appealing relation
between the CKM mixing angles
\be{appealing}
\frac{s_{13}}{s_{23}}=\frac{u}{d}\,s_{12}=0.11\div 0.15
\ee

However, there is a subtlety which we have avoided to discuss untill now.
Obviously, one 10-plet is not sufficient to obtain the realistic mass matrices.
The reason is that the symmetry
${\cal H}$ should transform the fermionic superfields in different way (in
order to assure the form of the heavy mass matrices (\ref{MassF}) by symmetry
reasons). Therefore, 10-plet is allowed to have at most 3 non-zero
Yukawa couplings in the eq. (\ref{WY}). The same is true for the $45_R$.
Obviously, three non-zero entries in the matrices $\Gamma$ and $G$ are not
enough - they will appear to be diagonal or degenerated. In order to supply in
(\ref{Mtot}) the off-diagonal entries $\hat{M}_{fL}$ and $\hat{M}_{fR}$ of the
non-trivial form, we need at least two 10's and two $45_R$'s:
$10_{1,2}$ and $45^R_{1,2}$, with different transformation
properties under ${\cal H}$, and all with non-zero VEVs. The Yukawa
superpotential is
\be{WYA}
W_Y=\Gamma^A_{ik} 10_A\, 16^f_i 16^F_k + G^B_{ik} 45^R_B\, 16^f_i
\overline{16}^F_k
\ee
Provided that both $45_R$-plets have non-zero VEVs on the $(1,1,3)$ component,
which also break ${\cal H}$ symmetry, we can change the basis and single out
one linear combination $45_R\propto V_{R1}45^R_1+V_{R2}45^R_2$, which
takes all the effective VEV $V_R=(V_{R1}^2+V_{R2}^2)^{1/2}$.
The other combination has vanishing VEV.

For the 10-plets the situation is more specific. In order not to affect
the gauge coupling crossing, only one combination of the two $\phi_1$ and
$\phi_2$, that are (1,2,2) components of $10_1$ and $10_2$, should remain
massless (in the SUSY limit), whereas other has to acquire $O(M_G)$ mass.
On the other hand, the (6,1,1) fragments $T_1$ and $T_2$ both should be
superheavy. Also, we do not want to pay {\em fine tuning} for this
doublet-triplet splitting. To achieve this, we suggest to modify
the Dimopoulos-Wilczek mechanism \cite{DiWi} in the following manner. Let us
introduce
yet another 10-plet, $10_0$, not necessarily coupled to fermions, and assume
that the ${\cal H}$ symmetry is designed so that allows only the following
couplings for 10's:
\be{WH}
\lambda 10_1 10_2 45_{BL} + \lambda_1 10_1 10_0 45^{R}_1 +
\lambda_2 10_2 10_0 45^{R}_2 + \lambda_0 10_0 10_0 54
\ee
Therefore, after substituting the relevant VEVs, mass matrices of the
$\phi$ and $T$ components of the 10's are
\be{M10}
\begin{array}{ccc}
 & {\begin{array}{ccc} 10_1\,\,\,\;\; & \;\,\,10_2 & \;\;\;\,\,\,\,\,10_0
\end{array}}\\ \vspace{2mm}
M_{\phi,T}~=~\begin{array}{c}
10_1 \\ 10_2 \\ 10_0   \end{array}\!\!\!\!\!&{\left(\begin{array}{ccc}
0 & \lambda\langle 45_{BL} \rangle & \lambda_1\langle 45^R_1 \rangle \\
-\lambda\langle 45_{BL} \rangle & 0 & \lambda_2\langle 45^R_2 \rangle \\
-\lambda_1\langle 45^R_1 \rangle & -\lambda_2\langle 45^R_2 \rangle &
\lambda_0 \langle 54 \rangle \end{array}
\right)} \end{array}
\ee
Considering $T$-components, we see that all three eigenstates are superheavy.
For the $(1,2,2)$ components we must substitute $\langle 45_{BL} \rangle
\rightarrow 0$ and $\langle 45^R_{1,2} \rangle\rightarrow V_{R1,2}$.
Then one linear combination $\phi\propto \lambda_2V_{R2}\phi_1 -
\lambda_1 V_{R1}\phi_2$ is massless, whereas the orthogonal combination is
superheavy. The VEV of $\phi$ (\ref{phi}), which arises after the
SUSY breaking, is shared between the ${\cal H}$ symmetry eigenstates $10_1$
and $10_2$.

The above consideration demonstrates how one could supply the realistic
structure for the fermion mass matrices. The form of the Matrices $\Gamma_A$
and $G_B$, $A,B=1,2,...$, are constrained by the ${\cal H}$ symmetry: no more
than three non-zero entries are allowed for each of them. However, for the
linear combinations $\phi$ and $45_R$ the eq. (\ref{WYA}) is effectively
reduced to (\ref{WY}), where the Yukawa matrices $\Gamma$ and $G$ can have
some non-trivial (e.g. Fritzsch) form.

Accidentally, the structure of the eq. (\ref{M10}) outwardly recembles the
one suggested by Babu and Barr \cite{BabuBarr} for the {\em strong}
suppression of the Higgsino mediated $d=5$ operators for the proton decay.
In our case, however, the {\em strong} suppression {\em a la'} Babu and Barr
does not occur, since both $10_1$ and $10_2$ are coupled to fermions.
Nevertheless, some {\em weak} suppression can be due to mixing of different
$T$-states from the $10_{0,1,2}$. On the other hand, our seesaw pattern
(\ref{Mtot}) in itself leads to the complete suppression of the dominant
($LLLL$-type) $d=5$ operators, and only much weaker $RRRR$-type ones
remain to be effective \cite{proton}. All this leaves us with the
chance to observe the proton decay at the level of present experimental bound.
It is worth to remark also, that in our scheme we can evaluate the branching
ratios of the different decay modes, since we are able to calculate
all mixing angles, including the ones for the charged leptons\footnote{
In fact, the existing calculations of the proton decay modes (see e.g.
\cite{Nath}) cannot be satisfactory, since they are performed within the
framework of the minimal SUSY $SU(5)$ model with obviously wrong mass
relations $d=e$ and $s=\mu$. }
This can be rather important for the testing of our {\em inverse hierarchy}
scheme, if the proton decay will be observed in the future.

Let us discuss now the neutrino mass and mixing pattern. Clearly, the eq.
(\ref{mf}) is valid also for the neutrino Dirac mass matrix, which has
the form of the eq. (\ref{form}) with $\eps_{\nu}=2\eps_e+3\eps_u$.
Then from the equations analogous to (\ref{eigen}) we obtain for the
neutrino Dirac mass eigenvalues at the GUT scale:
\be{Dirac}
\nu^D_1\approx \frac{2}{3}\,e\,,~~~~~\nu^D_2\approx \frac{3}{4}\,\mu\,,~~~~~
\nu^D_3\approx \frac{1}{4}\,\tau
\ee
These are drastically different from what is traditionally
expected from the $SO(10)$ model: for example, our $\nu^D_3$ is about
$200-300$ times less than in standard $SO(10)$ ($\nu^D_3=t$). We can also
evaluate the "mixing" angles, which diagonalize the matrices $\hat{m}_e$ and
$\hat{m}^D_{\nu}$, in the terms of the CKM angles:
$s^e_{12}\approx (3/4) s^D_{12}\approx (1/3)s_{12}=0.07$, etc. However,
all these do not transform into sharp predictions for the neutrino mass
and mixing pattern, since we do not know yet the Majorana mass matrix
of the right handed neutrinos.

For this purpose we have the Higgs 126-plet with VEV $v_R\sim M_G$.
However, there are too many
different possibilities to introduce its Yukawa couplings, all with different
implications for the neutrino mass end mixing pattern. For the demonstration,
we consider here one of the simplest possibilities. Let us suggest that the
126-plet interacts only with the $16^F$-plets: $\Lambda_{ij}126\,16^F_i16^F_j$,
where $\Lambda$ is a coupling constant matrix with $O(1)$ elements.  Then for
the Majorana mass matrix of the light neutrinos we immediatelly get
\be{Major}
\hat{m}^M_{\nu}=\frac{v^2}{2v_R}\Gamma \Lambda^{-1} \Gamma^T
\ee
This, in general, implies that all neutrinos have masses of the order of
$10^{-2}-10^{-3}$ eV, and their mixing angles are large,
which favours the adiabatic MSW solution to solar neutrino problem.
More precise information can be obtained
by constraining the form of the matrices $\Gamma$ and $\Lambda$ due to
${\cal H}$ symmetry properties.

\vspace{8mm}
{\bf 4. Discussion.}
\vspace{3mm}

Let us try to give some more philosophical shape to our considerations.
One could imagine that our SUSY $SO(10)\otimes {\cal H}$ theory is what
remains from the superstring after compactification. Obviously, such a
theory should be realized at some higher Kac-Moody level, since
we utilize the higher dimensional representations of $SO(10)$. In particular,
$k\geq 5$, if we use the $126$-dimensional representation for the symmetry
breaking and neutrino mass generation purposes \cite{Ibanez}. The fermionic
sector includes 5 zero modes of $16$-plets: $16^f_{1,2,3}$ and $16^F_{2,3}$,
and 2 zero modes of $\overline{16}$-plets: $\overline{16}^F_{2,3}$. We have
also included in game some non-zero modes like
$16^F_{0,1}+\overline{16}^F_{0,1}$,
with masses of the order of the compactification scale $M\sim$ few times
$10^{17}$ GeV. Taking seriously the coupling crossing phenomenon in MSSM,
we suggest that the breaking of $SO(10)\otimes {\cal H}$ symmetry down to
$SU(3)_c\otimes SU(2)_L\otimes U(1)_Y$ occurs at one step, at the scale
$M_G\sim 10^{16}$ GeV. All what remains below is just MSSM with three
quark-leptonic families that are fragments of the $16^f_{1,2,3}$, and one
couple of Higgs doublets, originated from certain effective combination of
the $\phi_A(1,2,2)$ components of the various $10_A$.
In order to render the couple of Higgs doublets massless in the exact SUSY
limit, we have used an intriguing modification of the Dimopoulos-Wilczek
mechanism.

We assumed that the generation of fermion masses occurs due to universal
seesaw mechanism. Once again we would like to stress that in seesaw picture
the ordinary light fermions $f$ are just the spectators of the phenomena
that determine the flavour structure. This structure arises in a sector of
the superheavy $F$ fermions and is transferred to the light ones at their
decoupling. Namely, the heaviest $F$ family $F_1$ is unsplit since it has
$SO(10)\otimes {\cal H}$ invariant mass of the order of $M$. The lighter ones
$F_2$ and $F_3$ get the masses of the order of $M_G$ and $M_G^2/M$
respectively, due to effective operators involving the
Higgs $45_X$ with successively increasing power, and are thereby split.
As a result, the $f$'s
mass matrices, given by a seesaw mixing with the $F$'s, have the
{\em inverse hierarchy} form.

These mass matrices, which in eq. (\ref{mf-1}) are displayed as
$\hat{m}_f^{-1}$ for the convenience reasons, reproduce the fermion spectrum
(see Fig. 1) and mixing pattern in a very economical way.\footnote{Clearly,
both the {\em decoupling} and {\em scaling} hypothesis of the Sect. 1
are naturally fulfilled in this way. In the limit when $\eps_u=\eps_d$
we have the {\em scaling}: $u:c:t=d:s:b$ and all CKM angles are vanishing.
The {\em decoupling} can be seen in the following way: by putting
$\eps^2_{u,d}$ to zero (as parametrically smaller values compared to
$\eps_{u,d}$), assuming also that the third family masses $t$ and $b$ are
fixed, we see that $u,d\rightarrow 0$ and at the same time $s_{12},s_{13}
\rightarrow 0$. At the next step, bu putting $\eps_{u,d}$ to zero, we
see that $c,s\rightarrow 0$ and also $s_{23}\rightarrow 0$. }
They differ only
due to different, in general {\em complex} expansion parameters $\eps_f$,
$f=u,d,e,\nu$ (including the neutrino Dirac mass matrix),
where $\eps_f\sim M_G/M\sim 10^{-1}-10^{-2}$. These parameters are related
through the $SO(10)$ symmetry properties (see eq. (\ref{eps2}), so only
two of them, say $\eps_d$ and $\eps_u$ are independent. Due to common
mass factor $m$, the first family plays a role of a {\em mass unification
point}, and the $e-u-d$ mass splitting is understood by the same
mechanism that enhances the Cabibbo angle up to the $O(\sqrt{\eps_d})$ value.
Other mixing angles naturally are in the proper range (see
eq. (\ref{CKM})).
We have obtained a number of interesting mass formulas, from which it
follows that $m_t=100-150$ GeV, $m_b=4-5$ GeV, $m_s=100-150$ Mev and
$m_u/m_d=0.5-0.7$. We did not utilize any particular supersymmetry breaking
mechanism, therefore we do not have some certain perdictions for the
parameters of MSSM. However, independently on the concrete mechanism,
we have rather interesting prediction $\tan\!\beta\approx 1$, which can be
immediately tested on the accelerators of the next generation \cite{kun}.


In fact, we did not suggest any concrete example of our misterious
symmetry ${\cal H}$, that should support the inverse hierarchy pattern
of fermion mass matrices, the modified Dimopoulos-Wilczek ansatz for
natural doublet-triplet splitting and, in the end \cite{end}, the needed
VEV pattern. In principle,
${\cal H}$ can contain some set of discrete or abelian (Peccei-Quinn type)
symmetries. Alternatively, one can try to utilize global or discrete
$R$-symmetries. I am convinced that to find the working example of a
${\cal H}$ symmetry is rather a
cumbersome but available task for the smart model-builder.

We find it amusing that the idea of inverse hierarchy, implemented in
SUSY $SO(10)$ theory in a natural way, can explain the key features of the
fermion mass spectrum and weak mixing. Notice, that in contrast to all known
predictive frameworks for fermion masses (see e.g. \cite{Fritzsch,Jarlskog}),
we did not exploit any particular texture - a horizontal structure that
suggests an existence of  certain "zeros" in mass matrices. Moreover,
it is clear that in our mass matrices there can be no "zeros" - this
immediately would bring us to wrong predictivity. However, a {\em clever}
horizontal structure would enhance a predictive power of our approach.
In paricular, one can expect
that the ${\cal H}$ symmetry will constrain also the Yukawa coupling matrices
$\Gamma$ and $G$ so that they will have certain pattern with
certain "zero" elements. These "zeros" will not be seen directly in the quark
and lepton mass matrices of the eq. (\ref{mf}), but they will manifest
themselves through certain relations between parameters of the eq. (\ref{Q})
which we have treated before as independent.
The example of the succesfull relation (\ref{appealing}) obtained by
suggesting the Fritzsch texture for $\Gamma$ and $G$
demonstrates that may be "zeros" are not placed directly in the mass matrices,
where they are generally looked while "stitching the Yukawa quilt".

\vspace{8mm}
\begin{flushright}
{\em "With a little help from my friends..."} \\
\end{flushright}


It is a pleasure to thank Z. Ajduk, S. Pokorski and other organizers of the
Warsaw meeting for their warm hospitality at Kazimierz. I am indebted to
my collaborator R. Rattazzi, together with whom the inverse hierarchy ansatz
was elaborated in a radiative scenario of the ref. \cite{rattazzi3}.
Also the useful discussions with R. Barbieri and S. Pokorski are
acknowledged. Many thanks are due to Ursula Miscili for her patience and
encouragement during my work on the manuscript.

\end{document}